\documentclass[prb,twocolumn,showpacs,preprintnumbers,amsmath,amssymb]{revtex4}

\usepackage{graphicx}% Include figure files
\usepackage{dcolumn}% Align table columns on decimal point
\usepackage{bm}% bold math

\usepackage{epsfig}
\usepackage{graphics}
\usepackage{color}
\usepackage{amsmath}

\usepackage{hyperref}
\hypersetup{
    colorlinks,%
    citecolor=blue,%
    filecolor=black,%
    linkcolor=blue,%
    urlcolor=black
}

\begin{document}

\title{Magnetic Phase Dependence of the Anomalous Hall Effect in Mn$_3$Sn Single Crystals}

\author{N. H. Sung$^{\star}$, F. Ronning, J. D. Thompson, E. D. Bauer }
\affiliation{Los Alamos National Laboratory, Los Alamos, NM 87545, USA
}
%\maketitle

\begin{abstract}

Thermodynamic and transport properties are reported on single crystals of the hexagonal antiferromagnet Mn$_3$Sn grown by the Sn flux technique.  Magnetization measurements reveal two magnetic phase transitions at $T_1$ = 275 K and $T_2$ = 200 K, 
%accompanied by an abrupt change of the Hall resistivity and a change in sign of the Hall coefficient at $T_1$ and at $T_2$, respectively, 
below the antiferromagnetic phase transition at $T_N$ $\approx$ 420 K. The Hall conductivity in zero magnetic field is suppressed dramatically from 4.7 $\Omega^{-1}$cm$^{-1}$ to near zero below $T_1$, coincident with the vanishing of the weak ferromagnetic moment. This illustrates that the large anomalous Hall effect (AHE) arising from the Berry curvature can be switched on and off by a subtle change in the symmetry of the magnetic structure near room temperature. 
%\textcolor{red}{ Furthermore, we demonstrate that the information stored in the AHE at high temperature can be partially preserved despite the change in magnetic structure at low temperature. }

\end{abstract}

\maketitle
\newpage

Conducting materials exhibit the Hall effect, a transverse voltage produced by longitudinal current flow in an orthogonally applied external magnetic field.  
The anomalous Hall effect (AHE), commonly associated with ferromagnets, is governed by extrinsic and intrinsic contributions.\cite{Nagaosa_2010}
The extrinsic contribution depends on details of the impurity scattering, while the intrinsic contribution, first considered by Karplus and Luttinger,\cite{PR1954} arises from a fictitious magnetic field due to Berry-phase curvature in momentum-space, with a strength determined by the spin-orbit coupling.\cite{Nagaosa_2010, DiXiao2010}  Experiments show a crossover from a dominant extrinsic effect on the AHE to a dominant intrinsic one with increasing longitudinal resistivity.\cite{YTokuraPRL2007}  
%Recently, an AHE has been observed in paramagnets, such as the pyrochlore Pr$_2$IrO$_7$, where the conduction electrons acquire a Berry-phase while moving in a non-coplanar (chiral) spin texture of Pr ions even in zero applied magnetic field.\cite{YoMachida2010}
%Here, geometrical frustration precludes long-range magnetic of the Pr magnetic moments, but instead generates a "2-in, 2-out" chiral spin liquid of Pr moments within the pyrochlore tetrahedron, which breaks time-reversal symmetry and generates an AHE, often called a topological Hall effect.\cite{Nagaosa_2010, Sci2001, PSSRRL2017, MingdaLi2016}

\begin{figure}
\centering
\includegraphics[width=0.4\textwidth]{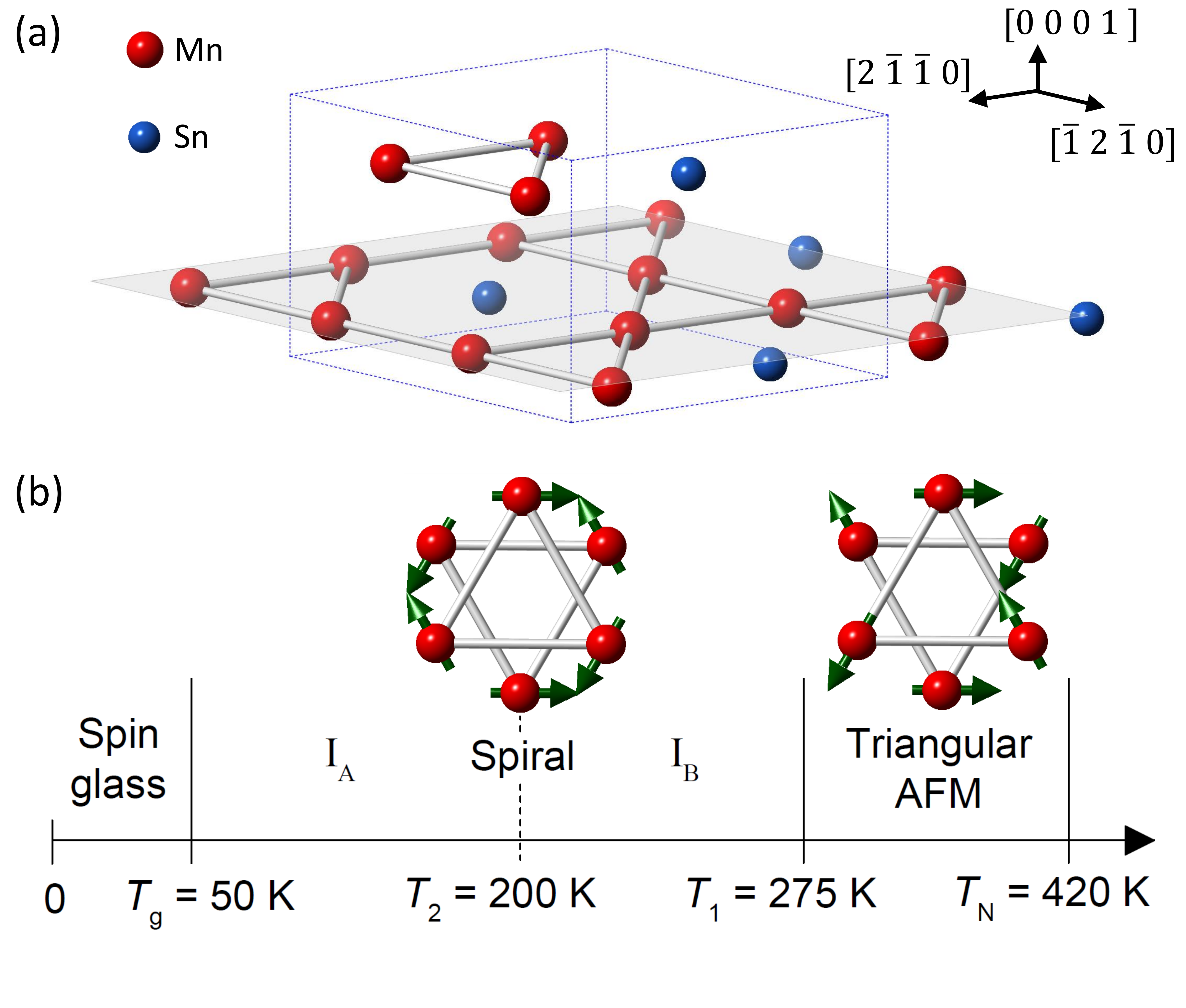}
\caption{
(a) Crystal structure of Mn$_3$Sn. Red and blue balls represent Mn and Sn atoms, respectively.
Dotted lines indicate the hexagonal unit cell and sticks show the kagome lattice in the (0 0 0 1) plane. 
(b) Schemetic magnetic phase diagram and spin structures of Mn$_3$Sn which indicate the spin glass, spiral, and triangular AFM states as a function of temperature.
The [1 1 $\bar{2}$ 0] direction and [0 0 0 1] direction are easy axes in the spiral spin states, I$_A$ and I$_B$, respectively.\cite{APL2015}
Green arrows indicate the suggested in-plane spin structures above and below $T_1$. Note the existence of a $C_{3z}$ symmetry below $T_1$ , which is broken in the inverse spin triangular phase above $T_1$. This enables a large AHE above $T_1$.
}\label{fig:fig04(Mn3Sn)}
\end{figure}

Antiferromagnets have recently received  attention for potential applications because of their insensitivity to perturbations, no stray fields, and fast spin dynamics that are required for data retention, high-density memory integration, and ultrafast data processing.\cite{NatureNanotech, arxiv1606, Sci2001} 
While an AHE is not realized in collinear antiferromagnets, recent theoretical and experimental investigations of 
%the Mn$_3$$X$ ($X$=Ga, Ge, Rh, Sn, Ir and Pt) 
chiral antiferromagnets reveal that a large AHE is possible, even in zero magnetic field, comparable in magnitude to that of ferromagnets.\cite{BinghaiYanPRB95, Nakatsuji2015, NakatsujiMn3Ge, SParkin2016}  
%The Mn$_3$$X$' ($X$'= Rh, Ir, and Pt) compounds crystallize in a cubic structure (space group $Pm$$\bar{3}$$m$), with the Mn atoms forming a Kagome lattice in the ($111$) plane.  Geometrical frustration causes the Mn moments to  order in an approximate "all-in", "all-out" triangular non-collinear antiferromagnetic structure, alternating in successive ($111$) planes; the exchange interactions are large, leading to magnetic order at high temperature, $T_N$=853 K, 960 K, and 473 K for Rh, Ir, and Pt, respectively.\cite{BinghaiYanPRB95} 
%Here, an AHE has been attributed to breaking a combined time-reversal and mirror-plane symmetry.\cite{CHerring1966,MacDonald} 
In the hexagonal Mn$_3$$X$ ($X$=Ga, Ge, Sn) systems (space group $P$6$_3$/${mmc}$, Fig. \ref{fig:fig04(Mn3Sn)} (a)), geometrical frustration in the Kagome lattice of Mn atoms within the $ab$-plane leads to non-collinear antiferromagnetic order with the moments aligned at 120$^{\circ}$, forming an inverse triangular spin structure.\cite{TomiyoshiJPSJ, PJBrown}  In the presence of spin orbit coupling, the symmetry of the structure permits an anomalous Hall contribution as well as a finite magnetization, although the relative magnitude of each term can not be determined by symmetry alone. 
Experimentally, the presence of a weak ferromagnetic moment (WFM)\cite{JWCable1993, Nagamiya1979, Nagamiya1982, Tomiyoshi1995, JMMM54, TomiyoshiJPSJ} enables the alignment of various domains with a small magnetic field, which otherwise could have opposite AHE contributions.  The small coercive field of this weak ferromagnetic state results in large changes to the Hall resistivity ($\approx 10 \mu \Omega $cm) with a small magnetic field ($H_c \approx$ 500 Oe) at room temperature.\cite{Nakatsuji2015} Large thermal Hall and Nernst signals in Mn$_3$Sn also arise due to a similar mechanism.\cite{arXiv161206128, Ikhlas2017}
These properties may make this material attractive for switches or data storage. 
 
A promising theoretical framework for estimating the magnitude of the intrinsic AHE based upon details of the band structure shows that the magnitude of the AHE depends sensitively on details of the magnetic and electronic structures.\cite{MacDonald, MTSuzuki_arxiv2016, Sci2001, YoMachida2010, CFelser, BinghaiYan2017, LeonBalents2017}
% The compounds have been also regarded as platforms for realizing topological Weyl semimetallic behavior in the chiral AFM systems. \cite{BinghaiYan2017, LeonBalents2017}
%but further work is needed to elucidate the relation between the magnetic structure, electronic structure, and the occurence and magnitude of the AHE in these Mn$_3$$X$ materials.  
Mn$_3$Sn is well-suited for such an investigation.  
Neutron diffraction studies on Mn$_3$Sn revealed the existence of the inverse triangular spin structure below $T_N$ $\approx$ 420 K.\cite{TomiyoshiJPSJ, PJBrown} 
With this magnetic structure, first principles calculations of the AHE from the Berry curvature are in good agreement with the experimentally determined magnitude of the AHE.\cite{BinghaiYanPRB95, Zhang} 
It is well known, however, that the magnetic structure of Mn$_3$Sn depends sensitively on the precise chemical composition and synthesis conditions.\cite{1972, 1975} A schemetic magnetic phase diagram is shown in Fig. \ref{fig:fig04(Mn3Sn)}. Most studies find a transition from the inverse triangular state to an incommensurate magnetic structure below $T_1$ in which the WFM is also suppressed. $T_1$ can occur as high as 275 K. 
A recent magnetic torque study observed two transitions $T_1$=275 K and $T_2$=200 K, which are argued to result from different easy axes in the two spiral magnetic phases.\cite{APL2015} 
At temperatures between 100 and 50 K an increase of the magnetization with decreasing temperature is reported to be due to the onset of spin-glass behavior.\cite{PJBrown,JMMM54} 
The transitions to an incommensurate magnetic structure do not occur in Czochralski grown samples, which are slightly Mn-rich.\cite{Nakatsuji2015}

Here, we report measurements of the physical properties of Mn$_3$Sn single crystals, synthesized by the self flux method, to examine the sensitive relation between the AHE and the magnetic structure.  
In particular, we find a large Hall signal ($\approx$ $-$1.5$\sim$$-$3 $\mu \Omega$cm) in the inverse triangular spin state, %down to the spiral state
$T_1 < T < T_N$, which is comparable to that found in other studies.\cite{Nakatsuji2015, SParkin2016} 
%below which it is small ($\approx\pm$0.02 $\mu \Omega$cm) with an apparent sign change below $T_2$ in the  second spin spiral state.\cite{APL2015}    
When the magnetic structure becomes an incommensurate spin spiral below $T_1$ = 275 K, the Hall resistivity drops by two orders of magnitude across a first order phase transition. We propose a change in magnetic structure that could account for the sudden vanishing of the AHE. 
%\textcolor{red}{Impressively, the system is able to retain some of its magnetic domain information through the transition.}
This trait adds to the functionality of large AHE effects in chemically tuned Mn$_3$Sn.

%\cite{JWCable1993, Nagamiya1979, Nagamiya1982, Tomiyoshi1995, JMMM54, TomiyoshiJPSJ} 
%Simultaneously, a significant intrinsic AHE is expected due to the tilting which is aligned with the ferromagnetic moment. 
%Consequently, the small coercive field of this weak ferromagnetic state allows one to observe large changes in the Hall resistivity ($\approx 10 \mu \Omega $cm) in a small magnetic field ($H_c \approx$ 500 Oe) at room temperature.\cite{Nakatsuji2015} 
%This property may make this material attractive for switches or data storage. 
%Moreover, recent thermal transport measurements provide evidence for large thermal Hall and Nerst signals in Mn$_3$Sn between $T$$_2$ and $T$$_N$.\cite{arXiv161206128}  

%\section{Experimental Methods}

\begin{figure}

\centering
\includegraphics[width=0.4\textwidth]{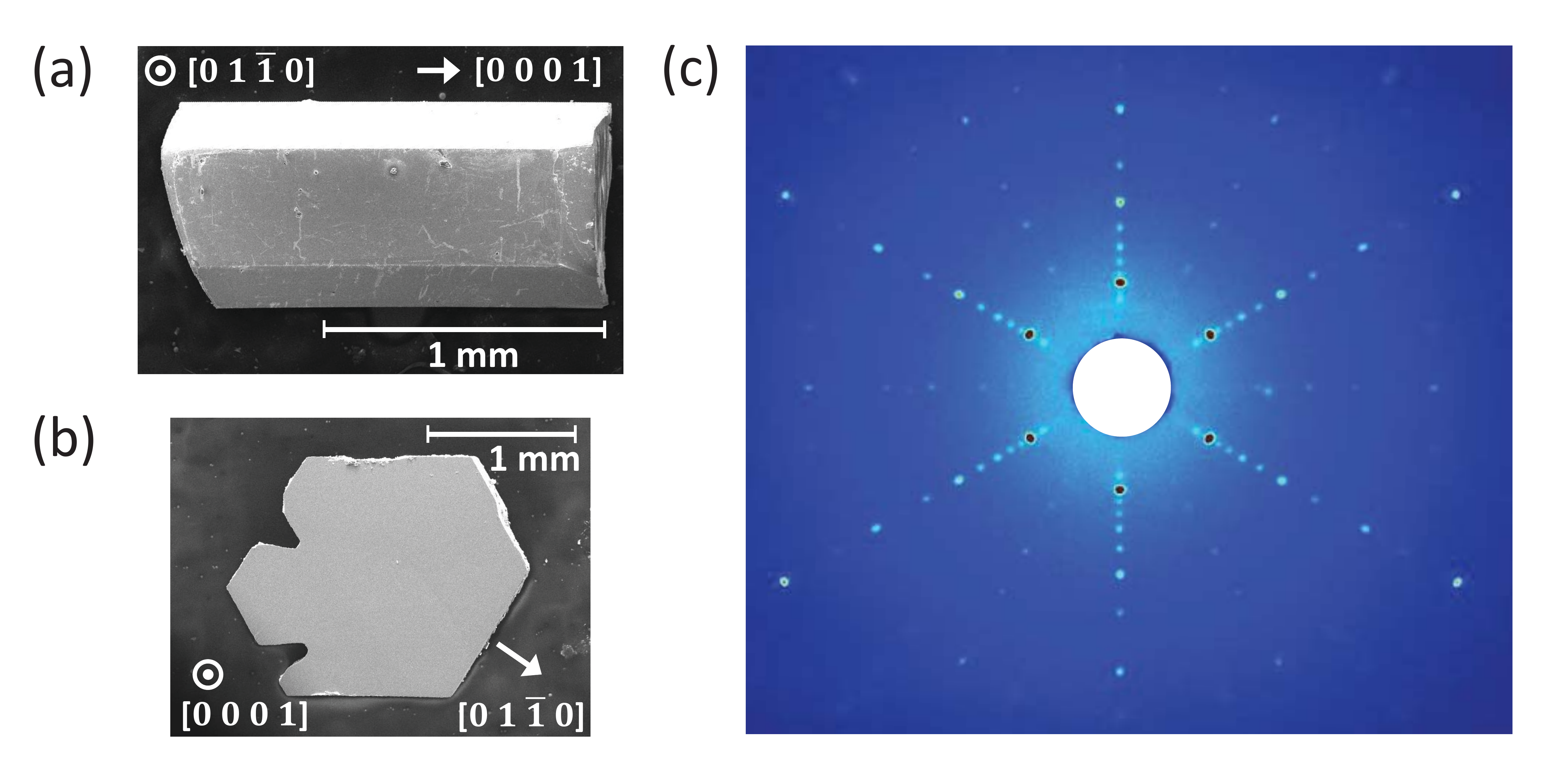}

\caption{
(Color online)
(a) and (b) show SEM images of Mn$_3$Sn single crystals with indications of crystallographic directions in the samples.
(c) Laue pattern of a polished surface showing the (0 0 0 1) plane.
}\label{fig:fig00(Mn3Sn)}

\end{figure}

Mn$_3$Sn single crystals were grown by the molten-metal flux growth method.\cite{Fisk1992}
Pieces of Mn (99.98\%) and Sn (99.999\%) were mixed with an atomic ratio of Mn : Sn = 7 : 3 and placed in an alumina crucible. 
The crucible was sealed in a silica tube under vacuum. 
The tube was heated  to 1000 $^\circ$C at a rate of 100 $^\circ$C/h,
%in a muffle furnace
maintained at 1000 $^\circ$C for 6 hours, and then cooled to 900 $^\circ$C at a rate of 1.25 $^\circ$C/h. 
At 900 $^\circ$C,  the silica tube was centrifuged to separate the  crystals from the flux using quartz wool as a filter. Mn$_3$Sn single crystals were obtained in the form of rods with a hexagonal cross section, with typical dimensions of $\sim$1 $\times$ 1 $\times$ 3 mm$^3$ (Figs. \ref{fig:fig00(Mn3Sn)} (a) and (b)).  
The crystals were characterized using Laue and powder x-ray diffractometers, scanning electron microscopy (SEM), and energy-dispersive x-ray spectroscopy (EDX; FEI Quanta 400 FEG-E-SEM) collected at 23 points at various points on the sample surface.
DC magnetization measurements of single crystals between 2.5 K and 380 K in magnetic fields up to $H$=50 kOe were performed in a Quantum Design Magnetic Property Measurement System (MPMS).  
Transport measurements were carried out in a Quantum Design Physical Property Measurement System (PPMS) between 1.8 K and 385 K and in magnetic fields up to 90 kOe. 
25-$\mu$m diameter Pt wires were spot welded to the sample surface  to make electrical contacts for electrical resistivity and Hall effect measurements. 
All Mn$_3$Sn samples were mechanically polished to remove any remaining flux on the surface before performing the measurements.

The temperature-dependent Hall resistivity ($\rho_{H}$) was measured at $H$ = 10 kOe and $H$ = 0 Oe, and defined as $\rho_{H}$($H$=10 kOe) = [$\rho_{H}$($H$=10 kOe)-$\rho_{H}$($H$=-10 kOe)]/2 and $\rho_{H}$($H$=0 Oe) = [$\rho_{H}$($H$$\rightarrow$+0 Oe)-$\rho_{H}$($H$$\rightarrow$-0 Oe)]/2, respectively. 
The notation +0 Oe (-0 Oe) refers to ramping the field to zero from positive (negative) fields ensuring no overshoot.
%To prevent a change in sign of magnetic field in ramping the field to zero, the field was decreased while maintaining a positive (or negative) field direction, notated as +0 Oe (or -0 Oe).
The Hall conductivity in zero field, $\sigma_H$($H$=0) is given by $\sigma_H$($H$=0) = -$\rho_{H}$($H$=0)/$\rho_{xx}^2$($H$=0) in the limit that the longitudinal electrical resistivity, $\rho_{xx}$, is much larger than $\rho_{H}$.
%\cite{Nakatsuji2015}
The field dependence of the topological AHE, $\rho^{A}_{H}$, is determined as $\rho^{A}_{H}$ = $\rho_{H}$ - $R_{0}B$ - $R_{s}\mu_{0}M$ where $R_0$, $R_s$ and $\mu_0$ are the normal and anomalous Hall coefficients and the permeability, respectively.
Here, $R_0$$B$ and $R_{s}\mu_{0}M$ indicate the ordinary Hall effect and the anomalous Hall effect due to magnetism in the compound, while $\rho^{A}_{H}$ is the topological Hall effect arising from spin texture due to the Berry phase curvature in momentum space.

%\section{Results and Discussion}

Mn$_3$Sn crystallizes in the hexagonal Ni$_3$Sn-type structure ($P$6$_3$/${mmc}$) as shown in Fig.  \ref{fig:fig04(Mn3Sn)} (a). 
The lattice parameters determined from the powder x-ray diffraction data were $a$ = 5.6763(4) {\AA} and $c$ = 4.5361(5) {\AA} (Fig. S1), comparable to the values found in the previous study.\cite{XRD}
The crystallographic directions of the samples are indicated in Figs. \ref{fig:fig00(Mn3Sn)} (a) and (b), which were determined by Laue diffraction (Fig. \ref{fig:fig00(Mn3Sn)} (c)).  
%The Laue patterns on a hexagonal cross section of the Mn$_3$Sn sample showed the (0 0 0 1) plane as displayed in Fig. \ref{fig:fig00(Mn3Sn)} (d). 
The chemical composition of the samples was determined to be Mn$_{2.97(1)}$Sn$_{1.03(1)}$ based on EDX measurements (Fig. S2), which is close to the ideal chemical composition, and henceforth, we refer to this sample as Mn$_3$Sn.

\begin{figure}

\centering
\includegraphics[width=0.45\textwidth]{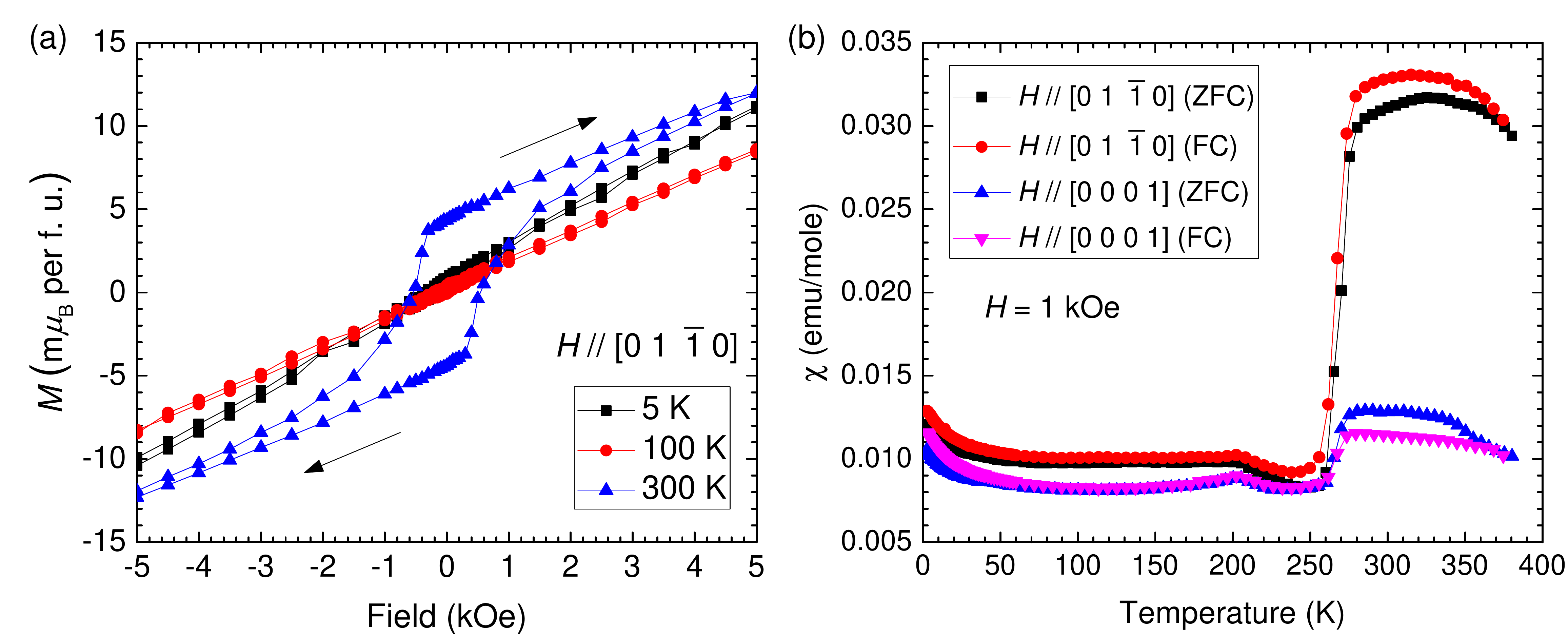}

\caption{ 
(Color online)
(a) Isothermal magnetization, $M$($H$), of a Mn$_3$Sn single crystal at $T$ = 5, 100, 300 K for $H$ $\parallel$ [0 1 $\bar{1}$ 0]. 
(b) Temperature dependent magnetization divided by field, $M$($T$)/$H$, of a Mn$_3$Sn single crystal with an applied field of $H$ = 1 kOe along $H$ $\parallel$ [0 1 $\bar{1}$ 0] and $H$ $\parallel$ [0 0 0 1] directions, in both ZFC and FC modes.
}\label{fig:fig01(Mn3Sn)}

\end{figure}

The isothermal magnetization, $M$($H$), of a Mn$_3$Sn single crystal was measured at $T$ = 5 K, 100 K, and 300 K in an applied magnetic field along $H$ $\parallel$ [0 1 $\bar{1}$ 0] (Fig. \ref{fig:fig01(Mn3Sn)} (a)). 
Weak ferromagnetism with hysteresis was observed at $T$ = 300 K, 
%which are similar to the previous results of Czochralski-grown samples, \cite{Nakatsuji2015, Tomiyoshi1995} 
but the phenomena disappeared at lower temperatures. 
With an applied magnetic field of $H$ = 5 kOe, the magnetic moment was $\sim$11, 8.5, and 12 m$\mu_B$/f.u. at $T$ = 5, 100, and 300 K, respectively.
Therefore, the different magnetic phases were distinguishable from the value of the magnetic moment and the existence of hystersis in the $M$($H$) curves. 	
Figure \ref{fig:fig01(Mn3Sn)} (b) shows the temperature dependence of the magnetization divided by field $M$($T$)/$H$, which for simplicity we label as $\chi$($T$), in  $H$ = 1 kOe along the $H$ $\parallel$ [0 1 $\bar{1}$ 0] and $H$ $\parallel$ [0 0 0 1] directions in a temperature range of 2.5 K $\leq$ $T$ $\leq$ 380 K. 
Two magnetic phase transitions at $T_1$ = 275 K and $T_2$ = 200 K were found, which correspond well to previous neutron diffraction and magnetization meausurements results.\cite{Nagamiya1979, Ohmori1987}
The change around $T_1$ is most prominent when the measurements were performed for $H$ $\parallel$ [0 1 $\bar{1}$ 0].
The phase transition at $T_2$ is possibly related to the helical structure reconstruction.\cite{APL2015} 
%The previous magnetization measurement also indicated the transition around $T^*$=210 K which is close to $T_2$ and a spontaneous magnetism was observed between $T^*$ and $T_N$$\approx$420 K,\cite{Ohmori1987} while in this study the WFM was observed only above $T_1$ and not in the temperature range  $T_2$ $\leq$ $T$ $\leq$ $T_1$.
Below $T_g$ $\approx$ 50 K, an upturn in $\chi$($T$) appeared with decreasing temperature  and the hystersis in $\chi$($T$) were observed  between zero field cooled (ZFC) and field cooled (FC) temperature sweeps, which indicates a spin-glass state, as previously reported. \cite{PRBglass2006} 
The changes in magnetization at $T_1$ and $T_2$ were not observed in $\chi$($T$) measurements of samples that were synthesized using the Bridgman method.\cite{PJBrown, JMMM54}
Moreover, both transitions were absent in measurements on single crystals synthesized by the Czochralski method using a 10\% excess of Mn to account for losses during the growth process; the chemical composition of the Czochralski single crystal was Mn$_{3.02}$Sn$_{0.98}$.\cite{Nakatsuji2015}
This further confirms that Mn$_3$Sn samples possess different magnetic structures depending on the chemical composition and/or growth conditions, as previously reported.\cite{1972, 1975}

\begin{figure}

\centering
\includegraphics[width=0.45\textwidth]{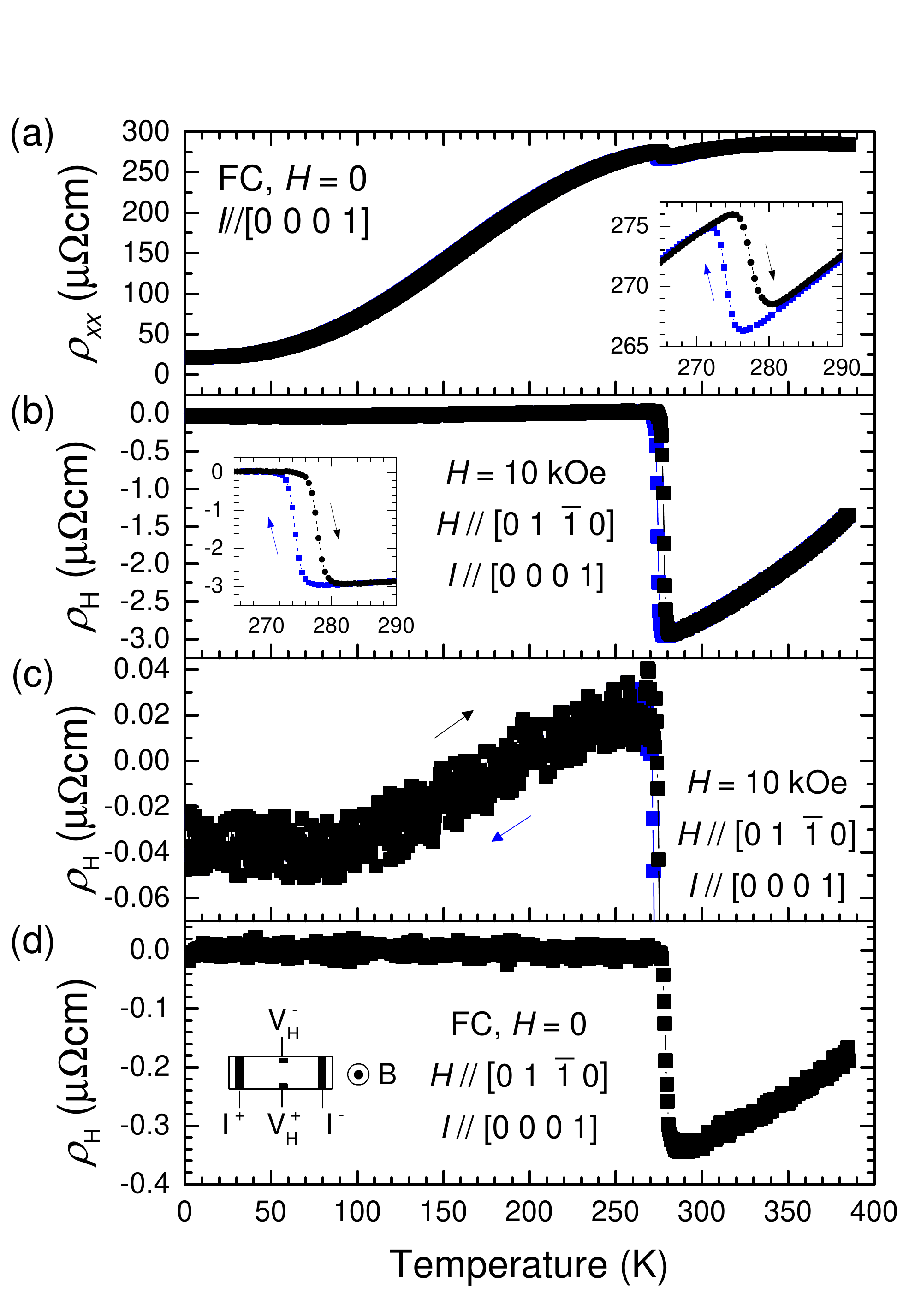}

\caption{
(Color online)
(a) Temperature-dependent zero-field electrical resistivity, $\rho_{xx}$($T$), of Mn$_3$Sn single crystals with the current along $I$ $\parallel$ [0 0 0 1].
%, obtained after the field cooling process in a magnetic field of $H$ = 10 kOe along $H$ $\parallel$ [0 1 $\bar{1}$ 0] direction
%Inset: expanded plot of $\rho_{xx}$($T$) around $T_1$. 
(b) Temperature-dependent Hall resistivity, $\rho_{H}$($T$), of a Mn$_3$Sn single crystal at $H$ = 10 kOe, $H$ $\parallel$ [0 1 $\bar{1}$ 0] and current $I$ $\parallel$ [0 0 0 1];
%The measurements were performed with heating (black circles) and cooling (blue squares) processes
insets: zoom of $\rho_{xx}$($T$) and $\rho_{H}$($T$) about $T_1$. 
(c) Expanded plot of $\rho_{H}$($T$). 
Black circles and blue squares indicate the obtained data as the temperature increases and decreases, repectively. 
(d) Temperature dependence of the zero-field Hall resistivity obtained after field cooling in a magnetic field of $H$ = 10 kOe along $H$ $\parallel$ [0 1 $\bar{1}$ 0] with $I$ $\parallel$ [0 0 0 1] and measuring upon warming. The inset shows the electrical contact configuration for the Hall measurements.
}\label{fig:fig02(Mn3Sn)}

\end{figure}

\begin{figure}

\centering
\includegraphics[width=0.4\textwidth]{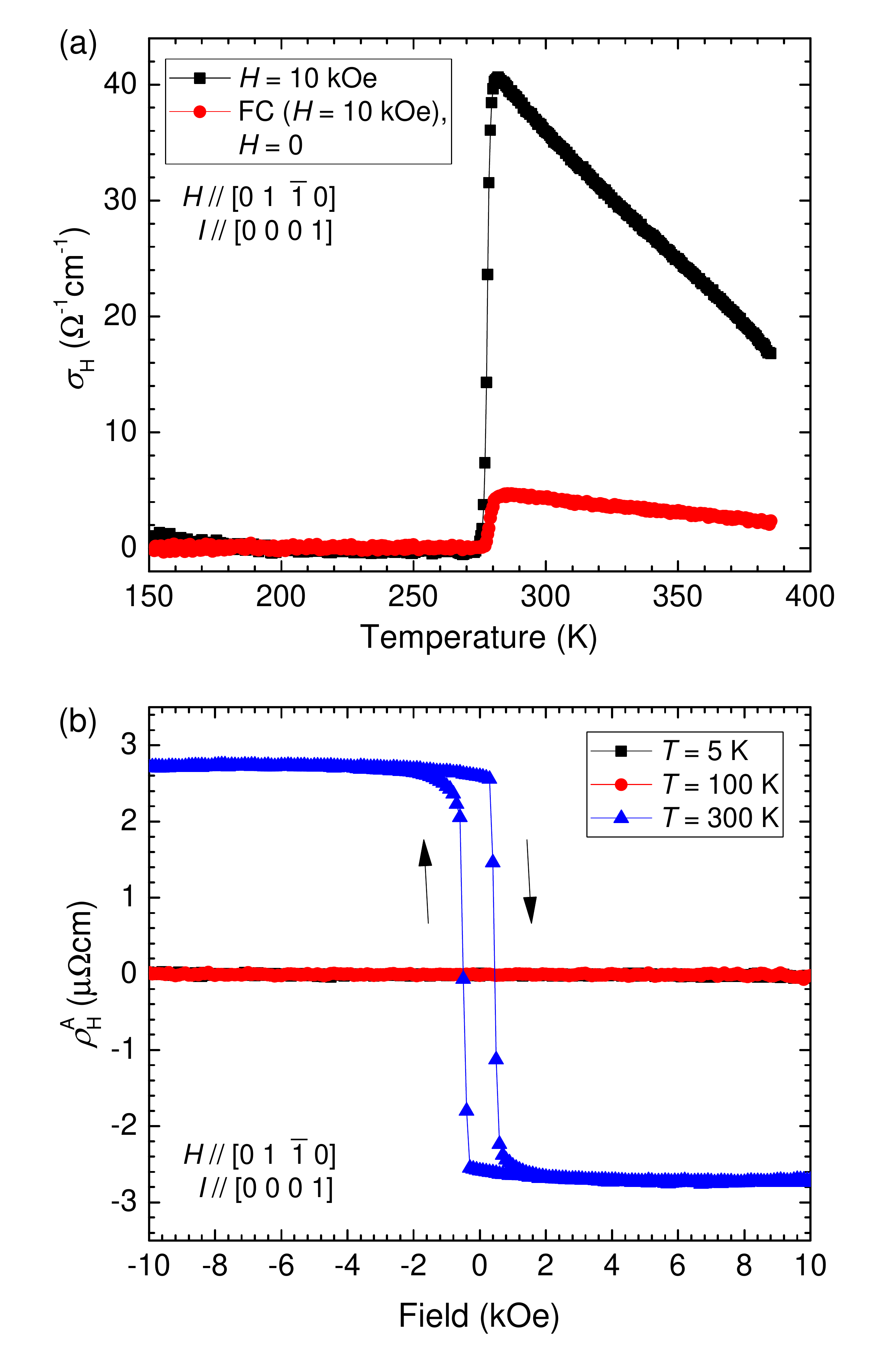}

\caption{
(Color online)
(a) Temperature dependence of the Hall conductivity in zero field, $\sigma_H$($H$ = 0), and in a field of 10 kOe, $\sigma_H$($H$ = 10 kOe) obtained from the resistivity data presented in Fig. 4.
(b) Magnetic field dependence of AHE, $\rho^{A}_{H}$, at $T$ = 5, 100, 300 K. The anomalous Hall contribution is zero below $T_1$. 
}\label{fig:fig03(Mn3Sn)}

\end{figure}

The electrical resistivity $\rho_{xx}$ with current $I$ applied along [0 0 0 1] as a function of temperature in zero field exhibits metallic behavior, as shown in Fig. \ref{fig:fig02(Mn3Sn)} (a), with a residual resistivity ratio ($\rho_{xx}^{385 K}$/$\rho_{xx}^{2 K}$) of 14.  
A kink in $\rho_{xx}$ was found at $T_1$ signaling the transition from the non-collinear AFM to a spiral magnetic state.\cite{JWCable1993} 
In the temperature-dependent Hall resistivity ($\rho_{H}$) measurement with an applied magnetic field of $H$ = 1 T along $H$ $\parallel$ [0 1 $\bar{1}$ 0] and the electrical current along $I$ $\parallel$ [0 0 0 1] (Fig. \ref{fig:fig02(Mn3Sn)} (b)), an abrupt change was also found at $T_1$.
A similar, albeit much slower, suppression of the AHE was also observed in ref. \cite{arXiv161206128}.
The insets of Figs. \ref{fig:fig02(Mn3Sn)} (a) and (b) show hysteretic behavior of $\rho_{xx}$ and $\rho_{H}$ vs. $T$ near $T_1$ upon  warming and cooling, which indicates a first-order phase transition at $T_1$.
A change in sign of $\rho_{H}$ was observed around $T_2$.
A slope change in $\rho_{H}$ around $T_g$, as shown in Fig. \ref{fig:fig02(Mn3Sn)} (c), is regarded to be related to the spin-glass state.\cite{PRBglass2006}
%\cite{APL2015, JWCable1993} 
%and probably related to the transition from the helical magnetic structure to the triangular magnetic structure. 
%A slope change and a change in sign of $\rho_{H}$ were observed around $T_g$ and $T_2$,  respectively, as shown in Fig. \ref{fig:fig02(Mn3Sn)}(c), and are related to the spin glass state around $T_g$ and also probably to the transition from the helical magnetic structure to the triangular magnetic structure at $T_2$, respectively. \cite{APL2015, JWCable1993, PRBglass2006} 
In contrast, Czochralski-grown samples show continuous $\rho_{xx}$ and $\rho_{H}$ curves without anomalies at the magnetic transitions $T_1$ and $T_2$.\cite{Nakatsuji2015}
Clearly, the AHE is extremely sensitive to the distinct magnetic structures in Mn$_3$Sn.
%, and the temperature regions of the phases  depends on growth conditions  and/or on the Mn:Sn ratio.

The Hall resistivity, $\rho_{H}$, was also measured at zero field ($H$=0) between 2 K $\leq$ $T$ $\leq$ 385 K with increasing temperature and a current along $I$ $\parallel$ [0 0 0 1]. 
The measurements were made after FC in an applied field of $H$ = 10 kOe along $H$ $\parallel$ [0 1 $\bar{1}$ 0] direction, then removing the field at the lowest temperature of 2 K (Fig. \ref{fig:fig02(Mn3Sn)} (d)). 
In contrast to $\rho_H$(10 kOe) shown in Fig. \ref{fig:fig02(Mn3Sn)} (b), $\rho_H$(0 kOe) below $T_1$ is strictly zero as one would expect for an ordinary metal in zero applied field. 
Above, $T_1$, $\rho_H$(0 kOe) has the same functional form as $\rho_H$(10 kOe), but with roughly an order of magnitude smaller amplitude than $\rho_H$(10 kOe). 
The origin of this reemergent Hall contribution in zero field requires further investigation.
Fig. \ref{fig:fig03(Mn3Sn)} (a) shows the temperature dependence of the Hall conductivity in zero and 10 kOe fields, $\sigma_H$($H$=0) and $\sigma_H$($H$=10 kOe), determined upon warming, where $\rho_{H}$ was extracted from Figs. \ref{fig:fig02(Mn3Sn)} (b) and \ref{fig:fig02(Mn3Sn)} (d), respectively.

In Fig. \ref{fig:fig03(Mn3Sn)} (b), the field dependence of the intrinsic AHE, $\rho^{A}_{H}$, at $T$ = 5, 100, and 300 K is displayed. 
To obtain $\rho^{A}_{H}$ the linear-in-field normal contribution $R_0 B$ was subtracted from the data. From the Hall data in Fig. \ref{fig:fig02(Mn3Sn)} (b) below $T_1$, the absolute value of $R_0$ is less than 4 $\times$ 10$^{-4}$ cm$^{3}$/C . At 300 K we subtract an extrinsic anomalous contribution with $R_s$ = -0.6 cm$^{3}$/C obtained from the linear dependence of $\rho_{xy}$ versus magnetization at 300 K (not shown). These values are comparable in magnitude to those obtained from Czochralski grown crystals,\cite{Nakatsuji2015} although the sign of $R_s$ is opposite in the two cases. We note that the values of $R_0$ and $R_s$ are negligible compared to the intrinsic anomalous Hall contribution above $T_1$.
The large values of $\rho^{A}_{H}$ observed above $T_1$ are comparable to ferromagnetic Fe or Ni,\cite{Nagaosa_2010, YTokuraPRL2007, Nakatsuji2015} but negligibly small values are found  below $T_1$. 
The results for $\rho^{A}_{H}$ at room temperature are qualitatively similar although differ slightly in magnitude to previous published results,\cite{Nakatsuji2015, SParkin2016} revealing that details of the electronic structure and/or scattering rate can influence the magnitude of the AHE in the inverse triangle magnetic state as might be expected.

Though details of the electronic structure determine the magnitude of the intrinsic AHE, the presence or absence of such a term is dictated by symmetry.\cite{MTSuzuki_arxiv2016} 
The inverse triangle magnetic state with magnetic ordering wave vector $\bf Q$=0 at high temperatures in Mn$_3$Sn breaks the many requisite symmetries to allow an intrinsic AHE contribution. 
Previous neutron scattering work on Mn$_3$Sn below $T_1$ have emphasized the incommensurate magnetic structure, which propagates along the $c$-axis.\cite{PJBrown, JWCableJAP1994} Naively, one would expect that an incommensurate magnetic structure would break additional symmetries, and hence it is surprising that the AHE vanishes below $T_1$. 
This suggests that the magnetic structure must fundamentally change to restore a symmetry below $T_1$ that was absent above $T_1$.
%, as opposed to simply additionally developing an incommensurate modulation of the structure. 
One such symmetry would be a three-fold rotation about the $c$-axis ($C_{3z}$). As shown in Fig. \ref{fig:fig04(Mn3Sn)}, this symmetry is broken above $T_1$, but is satisfied in the proposed in-plane magnetic structure below $T_1$. The presence of a $C_{3z}$ symmetry forces the Hall conductivity due to the Berry curvature for $H\perp c$ to be identically zero.\cite{MTSuzuki_arxiv2016}
Alternatively, the intrinsic AHE strongly depends sensitively on the orientation of the magnetic structure. Hence, it is conceivable that the incommensurate magnetic structure, could effectively average out the intrinsic AHE to a finite, but small, non-zero value.

Mn$_3$Sn single crystals synthesized by a simple self-flux method are found to possess a large intrinsic AHE in the so-called inverse triangle magnetic state comparable to that found in crystals grown by other techniques.\cite{Nakatsuji2015, SParkin2016}
Importantly, the different synthesis conditions leads to a sharp first order phase transition at 275 K to a new magnetic structure, which causes the intrinsic contribution to the AHE to abruptly vanish. The temperature at which these effects occur are sensitive to growth conditions and/or the Mn:Sn ratio. Consequently, this demonstrates an additional control parameter of the large anomalous Hall effect exhibited by antiferromagnetic Mn$_3$Sn, which could be exploited for technological applications.\\

\textbf{Supplementary Material}\\
See supplementary material for the powder x-ray diffraction and EDX measurements.

\acknowledgments
We thank Priscila Rosa, Hua Chen, Fengcheng Wu, and Ivar Martin for helpful discussions. 
Work at Los Alamos National Laboratory was supported by the Los Alamos LDRD program.
 N. H. Sung acknowledges a Director's Postdoctoral Fellowship also through the Los Alamos LDRD program. 
% performed under the auspices of the US Department of Energy, Office of Basic Energy Sciences, Division of Materials Sciences and Engineering.
The EDX measurements were performed at the Center for Integrated Nanotechnologies, an Office of Science User Facility operated for the U.S. Department of Energy (DOE) Office of Science. Los Alamos National Laboratory, an affirmative action equal opportunity employer, is operated by Los Alamos National Security, LLC, for the National Nuclear Security Administration of the U.S. Department of Energy under contract DE-AC52-06NA25396. 
\\

\noindent $^{\star}$ Corresponding author: nsung@lanl.gov

{}

\end{document}